\def\HA{{H$\alpha$}}
\def\HB{{H$\beta$}}

\def\kms{\:\rm{\,km\,s^{-1}}}

\def\LUM{\:{\rm ergs\:s^{-1}}}
\def\FLUX{\:{\rm ergs\:cm^{-2}\:s^{-1}}}

\def\VEL{\:{\rm km\:s^{-1}}}

\def\OiL{[\ion{O}{3}] $\lambda 6300$}
\def\OiiL{[\ion{O}{2}] $\lambda\lambda 3727, 3729$}
\def\OiiiL{[\ion{O}{3}] $\lambda\lambda 4959,5007$}
\def\SiiL{[\ion{S}{2}] $\lambda\lambda 6717, 6731$}
\def\NiiL{[\ion{N}{2}] $\lambda\lambda 6548, 6583$}
\def\oi{[\ion{O}{1}]}
\def\OiL{[\ion{O}{1}] $\lambda\lambda 6300, 6363$}
\def\HeiL{[\ion{He}{1}] $\lambda 5875$}
\def\AriiiL{[\ion{Ar}{3}] $\lambda 7135$}

\def\sii{[\ion{S}{2}]}

\def\oiii{[\ion{O}{3}]}

\def\hii{\ion{H}{2}}

\documentclass[12pt,preprint]{aastex}

\begin{document}


\newcommand{\MSOL}{\mbox{$\:M_{\sun}$}}  

\newcommand{\EXPN}[2]{\mbox{$#1\times 10^{#2}$}}
\newcommand{\EXPU}[3]{\mbox{\rm $#1 \times 10^{#2} \rm\:#3$}}  
\newcommand{\POW}[2]{\mbox{$\rm10^{#1}\rm\:#2$}}
\newcommand{\SING}[2]{#1$\thinspace \lambda $#2}
\newcommand{\MULT}[2]{#1$\thinspace \lambda \lambda $#2}
\newcommand{\CHINU}{\mbox{$\chi_{\nu}^2$}}
\newcommand{\vsini}{\mbox{$v\:\sin{(i)}$}}

\newcommand{\fuse}{{\it FUSE}}
\newcommand{\hst}{{\it HST}}
\newcommand{\iue}{{\it IUE}}
\newcommand{\euve}{{\it EUVE}}
\newcommand{\einstein}{{\it Einstein}}
\newcommand{\rosat}{{\it ROSAT}}
\newcommand{\chandra}{{\it Chandra}}
\newcommand{\xmm}{{\it XMM-Newton}}
\newcommand{\swift}{{\it Swift}}
\newcommand{\asca}{{\it ASCA}}
\newcommand{\galex}{{\it GALEX}}
\newcommand{\cxo}{CXO\,1337}


\shorttitle{SNRs in M33}
\shortauthors{Long \etal}

\title{
{Recovery of the Historical SN1957D in X-rays with Chandra\footnote{
Based on observations made with NASA's \chandra\ X-ray Observatory, 
the NASA/ESA Hubble Space Telescope,
the 6.5 meter Magellan Telescopes located at Las Campanas Observatory, 
and the Gemini Observatory.
NASA's \chandra\ Observatory is operated by Smithsonian Astrophysical Observatory 
under contract \# NAS83060 and the data were obtained through program GO1-12115.
The \hst\ observations were obtained at the Space Telescope Science Institute,
which is operated by the Association of Universities for Research in Astronomy, Inc. (AURA),
under NASA contract NAS 5-26555.
The new \hst\ observations were obtained through programs \# GO-12513 and GO-12683. Data 
in the \hst\ archive from program GO-11360 was also used.
The ground-based observations were obtained through NOAO
which is operated by Association of Universities for Research in Astronomy, Inc. 
for the National Science Foundation.}}
}

\author{
Knox S. Long\altaffilmark{2, 9},
William P. Blair\altaffilmark{3, 9},
L. E. H. Godfrey\altaffilmark{4},
K. D. Kuntz\altaffilmark{3}, \\
Paul P. Plucinsky\altaffilmark{5}, 
Roberto Soria\altaffilmark{4},
Christopher J. Stockdale\altaffilmark{6,7},\\
Bradley C. Whitmore\altaffilmark{2}, and
P. Frank Winkler\altaffilmark{8,9}
}

\altaffiltext{2}{Space Telescope Science Institute, 3700 San Martin Drive, 
Baltimore, MD, 21218;  long@stsci.edu, whitmore@stsci.edu}
\altaffiltext{3}{The Henry A. Rowland Department of Physics and Astronomy, 
Johns Hopkins University, 3400 N. Charles Street, Baltimore, MD, 21218; 
kuntz@pha.jhu.edu, wpb@pha.jhu.edu}
\altaffiltext{4}{Curtin Institute of Radio Astronomy, Curtin University, 
1 Turner Avenue, Bentley WA6102, Australia; leith.godfrey@icrar.org, roberto.soria@icrar.org}
\altaffiltext{5}{Harvard-Smithsonian Center for Astrophysics, 60 Garden Street, 
Cambridge, MA 02138; plucinsky@cfa.harvard.edu}
\altaffiltext{6}{The Homer L. Dodge Department of Physics \& Astronomy,
The University of Oklahoma,
440 W. Brooks St.,
Norman, OK 73019; Christopher.Stockdale@marquette.edu} 
\altaffiltext{7}{The Australian Astronomical Observatory,
PO Box 296,
Epping NSW 1710} 
\altaffiltext{8}{Department of Physics, Middlebury College, Middlebury, VT, 05753; 
winkler@middlebury.edu}
\altaffiltext{9}{Visiting Astronomer, Gemini Observatory, La Serena, Chile}

\begin{abstract}

SN1957D, located in one of the spiral arms of M83, is one of the small number of extragalactic supernovae that has remained
detectable at radio and optical wavelengths during the decades after its explosion.  Here we report the first detection of
SN1957D in X-rays, as part of a 729 ks observation of M83 with \chandra.  The X-ray luminosity (0.3 - 8 keV) is \EXPU{1.7^{+2.4}_{-0.3}}{37}{\LUM}.
The spectrum is hard and highly self-absorbed compared to most sources in M83 and to other young supernova remnants,
suggesting that the system is dominated at X-ray wavelengths by an energetic pulsar and its pulsar wind nebula.
The high column density may be due to absorption within the SN ejecta.
\hst\ WFC3 images resolve the supernova remnant from
the surrounding emission and the local star field. Photometry of stars around SN1957D,
using WFC3 images, indicates an age of less than 10$^7$ years and
a main sequence turnoff mass more than 17 \MSOL.  New spectra obtained with Gemini-South show
that the optical spectrum  continues to be dominated by broad \oiii\ emission
lines, the signature of fast-moving SN ejecta.
The width of the broad lines has remained about 2700 $\VEL$ (FWHM).
The \oiii\ flux dropped precipitously between 1989 and 1991, but
continued monitoring shows the flux has been almost constant since.
In contrast, radio observations over the period 1990-2011
show a decline rate $S_{\nu} \sim t^{-4.0}$, far steeper than the rate observed
earlier, suggesting that the primary shock has
overrun the edge of a pre-SN wind.

Subject Headings: galaxies: individual (M83) -- galaxies: ISM -- radio continuum: galaxies -- supernova remnants -- X-rays: general -- X-rays: individual (SN 1957D, M83) -- supernovae: individual (SN 1957D)

\end{abstract}

\section{Introduction \label{sec_intro}}

M83 (NGC 5236) is a nearby grand-design spiral with prolific star formation that has been the host to six historical supernovae (SNe), including SN1957D.  Relatively little is known about the SN1957D.   It was observed at a photographic magnitude m$_{pg} \lesssim 15$ \citep{kowal71}, which corresponds to an absolute magnitude of -13.3 using an assumed distance to M83 of 4.61 Mpc \citep{saha06}.  It was probably well past peak brightness at the time when it was observed.  The SN is located at the inner edge of a prominent northern spiral arm about 3 kpc from the nucleus of the galaxy.   No spectra during outburst are known, and so we do not know the SN type from original observations.  However, because of its proximity to active star formation, a core-collapse SN of some type is strongly suspected.


SN1957D was first recovered as a young supernova remnant (SNR)  at radio wavelengths in 1981, 24 years after the SN explosion, by \cite{cowan85}.   A nonthermal source, it had  a monochromatic radio luminosity similar to that of Cas A at that time.  Its flux  declined over the next few years, as do most old SNe/young SNRs that are detected in this age range \citep{stockdale06}.  Of the six historical SNe in M83, three -- SN1923A, 1950B, and 1957D -- remained detectable as radio sources years after the SN event.  SN1983N, classified as a type Ib SN, was detected at radio wavelengths just after the explosion, but faded below detectability very quickly \citep{cowan85,maddox06}. The two other historical SNe, SN1945B and SN1968L, have not been detected in the radio.  SN1968L lies is in a region of bright, diffuse radio emission near the nucleus, making detection of a radio SNR much more difficult than the others with arcsecond scale angular resolution.\footnote{SN1968L has been tentatively identified optically in \hst/WFC3 images \citep{dopita10}.}

SN1957D was recovered at optical wavelengths in 1987, 30 years after the SN explosion, by \cite{long89} as an \oiii\ emission-line source. By 1991, its flux declined by a factor of five when it was reobserved by \cite{long92}.  Spectra obtained during this period showed emission in \OiiL, \OiiiL\ and \OiL\ but little else.   Line widths were 2400-2900 $\VEL$ (FWHM) when the signal-to-noise ratio was high enough to measure the line width  \citep{long89, turatto89, long92}.   The spectra resemble those of other young SNRs thought to have arisen from core-collapse SNe, such as  Cas A and G292+1.8 in the Galaxy and  N132D, and E0102-72.3 in the Magellanic Clouds.  Emission in these O-rich SNRs is usually interpreted as arising from shock-heated knots of metal-enriched ejecta from the explosion of a massive star \citep{southerland95}.

SN1957D has not been detected previously at X-ray wavelengths, and in particular it was not detected by \cite{soria03} who analyzed a 50 ks observation of M83 obtained with \chandra\ in 2000.  However, it is clearly detected as a very hard X-ray point source in new, much deeper,  \chandra\ observations with a total exposure time of 729 ks.  This detection, along with new radio and optical observations of the SNR and its local environment, are the subject of this report.   In the next three sections, we describe the results of our analysis of the observations at X-ray, radio, and optical wavelengths.  Then in Section \ref{sec_discussion}, we compare SN1957D to other very young SNRs in order to obtain an overall picture of the SNR.  Finally, in Section \ref{sec_summary}, we summarize the main results.

\section{X-ray Observations -- Reductions and Analysis \label{sec_xray}}

The new \chandra\ observations of M83 were obtained in ten sections, starting in late 2010 and continuing throughout 2011 as part of a large program (12620596; Long P.I.) to characterize SNRs, point sources and diffuse emission in M83 to a limiting luminosity of about \EXPU{5}{35}{\LUM} (for spectral parameters typical of X-ray binaries).  The combined exposure totals 729 ks.  A detailed observing log appears in Table 1 of \cite{soria12}. We used the back-illuminated S3 chip in order to maximize the response at soft X-ray energies and we obtained the data in the ``very faint'' mode to optimize background subtraction.  We have used  tools within {\small CIAO} Version 4.4 \citep{fruscione06} to reduce the data, especially to create response and area corrected ancillary files needed for spectral fitting and we have used {\small XSPEC} Version 12 \citep{arnaud96} to characterize the X-ray spectrum.  

A 3-color X-ray image of the region of M83 containing SN1957D is shown in the upper left panel of Fig.\ \ref{fig_overview}. For comparison, an optical emission line image obtained at the 6.5m Magellan telescope at Las Campanas Observatory and optical continuum and emission line images obtained with  \hst\ are also shown.  There is a hard (blue) X-ray source coincident with an \oiii-dominated, point-like emission nebula that we identify as the young SNR from SN1957D (see following sections).  The SNR is also detected as a non-thermal radio source with ATCA (see Section \ref{sec_radio}). 

The uncertainty in the absolute astrometry of pipeline-processed \chandra\  images  is 0\farcs6 (90\% confidence) within 3\arcmin\ of the aim point.\footnote{See the analysis provided by the CXC at http://cxc.harvard.edu/cal/ASPECT/celmon/}  The statistical uncertainty in the position of the source simply due to the size of the point spread function and number of X-ray events detected is of order 0\farcs1 (1$\sigma$).   Our initial position estimate for the X-ray source identified at SN1957D was about 0\farcs3  from the ATCA radio position for SN1957D, so within the errors the positions are identical.  However, the absolute astrometry of the \chandra\ images can be improved by using the ATCA radio data.  Of the 65 sources observed with ATCA at 5 GHz, 25 lie within 1\arcsec\ of sources in our preliminary version of the M83 X-ray point source catalog.  If we exclude SN1957D and correct  the astrometry of the \chandra\ positions by the average position offset of the 24 remaining sources, the difference between the corrected X-ray source  and  radio position of SN1957D is 0\farcs14, is very close to the  statistical error.  The corrected (J2000) position for the object we identify with SN1957D is:

$${
\rm X\!\!-\!\!ray :  13^{h}37^{m}03.584^{s};  -29\degr49\arcmin40.91\farcs 
}$$


To obtain the X-ray spectrum of SN1957D, we defined a circular source region with radius 2\farcs5, and a background annulus between radii 3\arcsec\ and 6\farcs5. We extracted and combined source and background spectra from each observation, together with their response and ancillary response matrices.  We have fit the spectrum, which is shown in Fig.\ \ref{fig_xray_spectrum} and contains 140 counts after background subtraction,  to power-law and thermal models.  Due to the limited number of counts, we have used c-statistics rather than  $\chi^2$ statistics in the model fitting.  Fixing foreground Galactic absorption at \EXPU{4}{20}{cm^{-2}} \citep{kalberla05}, we find a power law with a photon power law index  $\gamma$ of $1.4^{+1.0}_{-0.8}$ and $N_H$(M83) of \EXPU{2.0^{+1.6}_{-1.1}}{22}{cm^{-2}}.\footnote{All of the errors quoted here represent 90\% confidence intervals.}  The derived 0.3-8 keV luminosity is \EXPU{1.7^{+2.4}_{-0.3}}{37}{\LUM}.  The spectrum can also be modeled equally well as with a thermal model, but in this case we  obtain only a lower limit of 10 keV to the temperature.  If X-ray emission from SN1957D is from a thermal plasma with this effective temperature, we would not expect to see lines in spectra with so few counts.

We searched for evidence of a source at the position of M83 in archival \chandra\ data obtained in 2000 and 2001 with a total exposure of 60,800 s.  The source was not detected although there was a slight excess at the known position of the SN. We estimate the 0.3-8 keV count rate to have been \EXPU{9.4\pm7.9}{-5}{counts~s^{-1}}, corresponding to a 3 $\sigma$ upper limit of \EXPU{3.3}{-4}{counts~s^{-1}}.  In 2010, the count rate was \EXPU{1.8\pm0.2}{-4} {counts~s^{-1}}. As a result, we cannot exlude the possibility that the source was as bright in the archival data as it is today, but we can rule out the possibility that it was much brighter in 2000-2001.

\section{Radio Observations -- Reductions and Analysis \label{sec_radio}}

M83 was observed with the Australia Telescope Compact Array (ATCA) in the 6A configuration on 2011 April 28-30\@. On each of the three days, a full 12-hour synthesis was obtained, simultaneously recording 2 GHz bandwidths centered on 5.5 GHz and 9 GHz in all four polarization products. Regular scans on the nearby phase calibrator PKS 1313-333 were scheduled throughout the observations, as well as scans on the ATCA primary flux calibrator PKS 1934-638. Standard calibration and editing procedures were carried out using the MIRIAD data analysis package \citep{sault95}. Following calibration, the data were exported for imaging, self-calibration and primary beam correction using the Common Astronomy Software Applications (CASA) package \citep{mcmullin07}. 
The 5.5 GHz data were imaged with uniform weighting and had angular resolution 2.52\arcsec x 1.06\arcsec\ at a position angle of 179\degr. The 9 GHz map was imaged with Briggs weighting (Robust = 0) and convolved to match the 5.5 GHz resolution. Coordinates were obtained by fitting an elliptical Gaussian to the pixels with values greater than half the peak, using the CASA task IMFIT.  We detected about 64 and 23 sources at 5.5 and 9 GHz respectively, including, at both frequencies, SN1957D.  A complete discussion of the ATCA radio observations of M83 will appear elsewhere.  


SN1957D is unresolved in both the 5.5 GHz and 9 GHz images and has a measured position (J2000) of 
$${
\rm Radio: 13^{h}37^{m}03.586^{s}\pm 0.003^{s};  -29\degr49\arcmin41.0\arcsec \pm 0.1\farcs
}$$

Flux densities for SN1957D were obtained at each frequency using an elliptical aperture centered on SN1957D, with major and minor axes equal to twice the restoring beam FWHM and rotated to the same position angle as the restoring beam. 
The flux densities are 230$\pm$10 $\mu$Jy and 190$\pm$10 $\mu$Jy at 5.5 GHz and 9 GHz respectively, giving a spectral index $\alpha = -0.4\pm$0.1 ($F_\nu \propto \nu^{\alpha}$). This spectral index is more typical of SNRs in which the radio emission is dominated by shocks than those in which the radio emission is dominated by a pulsar \citep{green09}.  There are no previous radio measurements of the flux from SN1957D at 9 GHz, and therefore we cannot directly compare radio spectral indices from these observations to those made by others at 6 and 20 cm, the most recent measurement of which is $\alpha = -0.26 \pm 0.14$ based on data taken in 1998 \citep{maddox06, stockdale06}.

The time evolution of the radio flux from SN1957D is shown in Fig.\ \ref{fig_time}.  A broken power law of the form 
\begin{eqnarray}
F_\nu &\propto& t^{-\delta_1} \qquad t < t_{\rm break}    \nonumber \\
F_\nu &\propto& t^{-\delta_2} \qquad  \nonumber t > t_{\rm break} \nonumber
\end{eqnarray}
reproduces the data, and captures the fact that the rate of declined has steepened over time. The power-law slope after the break is well constrained ($\delta_2 = 3.97^{+0.17}_{-0.20}$), but the location of the break ($t_{\rm break} < 33$ years since the SN explosion), and the power-law slope before the break ($\delta_1 < 1.5$) are not as well constrained. In Fig.\ \ref{fig_time} we have assumed $t_{\rm break} = 33$ years since the SN explosion --- the upper limit for $t_{\rm break}$. However, we might expect that the break is coincident with the rapid drop in \oiii\ flux seen in the period 1988-1989 \citep[][see also Section \ref{sec_optical} and Fig.\ \ref{fig_time}, bottom panel]{long92}, in which case the initial decline would be much flatter than the model fit in Fig.\ \ref{fig_time}, with slope of approximately $\delta_1 \sim 0.7$. 

\section{Optical Observations -- Reductions and Analysis \label{sec_optical}}

\subsection{Imaging Observations}

\cite{long92} discussed the early time-evolution of the \oiii\ emission from SN1957D,  noting that the flux had declined by a factor of five between 1987 and 1991.  We have continued to obtain narrow-band \oiii\ images  of the SN over the years to monitor its evolution.  An observing log for the various datasets we have available  is presented in Table \ref{table_optical}.   We have reduced all of the ground-based data  using standard IRAF\footnote{IRAF is distributed by the National Optical Astronomy Observatory, which is operated by AURA, Inc., under cooperative agreement with the National Science Foundation} procedures,  including flux calibration based on observations of standard stars from the list of \citet{hamuy92} and the known transmission functions for the filters.  We subtracted scaled continuum images from the emission-line ones and carried out conventional aperture photometry to measure the fluxes.  In the upper right panel of Fig.\ \ref{fig_overview}, we show a small portion of one of these data sets, from the 6.5m Magellan/IMACS continuum-subtracted emission-line images.  

The region of M83 containing SN1957D was observed with the \hst\ WFC3 in 2010 March as part of the Early Release Science program after Servicing Mission 4 {see Table \ref{table_hst}).    We obtained both narrow-band emission-line and broadband continuum data from the Mikulski Archive for Space Telescopes (MAST) at STScI and, starting with data reduced through the standard WFC3 pipeline, used MultiDrizzle \citep{multidrizzle} to combine the various exposures and put all the images on a common coordinate system.\footnote{These are the same data used by \citet{milisavljevic12}, but we have done the reductions and photometry independently.}  
A scaled version of the F547M (narrow V) continuum image was used to remove most of the stellar emission from the narrow-band images to produce pure emission-line images in H$\alpha$, \sii, and \oiii. These data were used to produce the lower right panel in Fig.\ \ref{fig_overview}, which is directly comparable to the Magellan data presented above.  The \hst\ data clearly resolve the point-like \oiii\ SNR emission from the surrounding more diffuse gas, and in particular a faint arc of H$\alpha$ emission to the south and east of the SNR position.   We have included our photometric measurements for the \oiii\ 5007 \AA\ emission along with the ground-based measurements in Table \ref{table_optical}.

The \hst\ broadband images of the region are also very interesting. In the lower left panel of Fig.\ \ref{fig_overview}, we show an aligned 3-color representation of the \hst\ U-B-narrow V data, which demonstrates the presence of a very blue association of stars just to the northeast of the SNR position.    Furthermore, careful alignment of the continuum and emission-line images shows that exactly at the position of the SNR  is a remaining star, whose color is somewhat redder than most of the other stars in the adjacent association.  This figure also shows that the extinction in the region is patchy and variable,  extending right to the edge of the association on the west side, and possibly overlying the SNR position itself.  We do not know if this star is associated with the SN precursor (e.g. a possible companion star) or whether it is a chance alignment along the line of sight, but its position is coincident with the SNR position within the measurement error.  The association will be discussed further below.

The precision of the flux values in Table \ref{table_optical} is limited primarily by our ability to distinguish SN1957D from the nearby \hii-region emission, and to accurately measure the underlying background.  Both problems are, of course, worse under conditions of poor seeing, but even with the \hst\ WFC3 data there is still some background contamination.  In order to minimize the confusion problem, we have used minimum aperture sizes (diameter $\approx$ FWHM) for the SN1957D photometry and have made aperture corrections derived from data on several isolated bright stars in each image.  We have also experimented with different  choices for background regions, and include variation from these, in addition to purely statistical errors, in the uncertainty estimates for Table \ref{table_optical}.    As noted in the table, there are minor differences in the pass bands for the different \oiii\ filters used for different observations, but all pass most of the broad 5007 \AA\ emission, and exclude most of the 4959 \AA\ line.  All the flux values also include the narrow component of the 5007 \AA\ line, since there is no way to distinguish between broad and narrow components in our images.

The time evolution of the \oiii\ flux derived from the above observations is shown in the bottom panel of Fig.\ \ref{fig_time}.   All of the measurements are consistent with an \oiii\ flux of about \POW{-15}{\FLUX} between 2002 and 2011, about 2.5 times smaller than the flux measured in 1991.  Clearly the rapid decline in the flux seen in the earlier data has halted.  Unfortunately we do not have any images between 1991 and 2002, but interpolating the existing data, it appears consistent that the rapid decline had essentially leveled-out by the early 1990's.

\subsection{Spectroscopic Observations}

As part of a Gemini observing program (\# 2011A-0436, Winkler PI) to obtain spectra of SNRs in M83, we obtained a spectrum of SN1957D using the Gemini Multi-Object Spectrograph (GMOS) on the 8.2m Gemini South telescope on 2011 April 8 (UT).  Our observations consisted of  three 40-minute exposures using the 600 line blue grating (G5323) in multi-slit mode.  For SN1957D, the custom slit was 1\farcs25 wide and 10\arcsec\ long, oriented E-W.  The spectrum,  extracted  from a 1.5\arcsec\ region along the slit and covering  the wavelength range 4700 - 7300 \AA, is shown in Fig.\ \ref{fig_optical_spectrum}.  The original two-dimensional spectrum has a blue continuum coincident with SN1957D but extending extending well beyond it to the east, which we attribute to the blue stellar association shown in Fig.~\ref{fig_overview} (lower left panel); for clarity we have fitted and subtracted that from the spectrum in Fig.\ \ref{fig_optical_spectrum}.
The spectrum shows broad lines from \OiiiL, \OiL,   \SiiL, and \AriiiL, as well as narrow lines from \HA, \HB, \HeiL, \OiiiL, \NiiL,   \SiiL,  and \AriiiL, as listed in Table 3.  
 Simultaneous fits for the broad components of the \oiii, \oi, and \sii\ lines gives a velocity width  of $2730 \pm 100 \kms$ (FWHM).
 From the \hst\ and ground-based imagery available, we conclude that some of the narrow-line emission probably arises from the near-coincident diffuse  emission located $\lesssim 1\arcsec$\ to the east and south of SN1957D,   since the two cannot be fully resolved at ground-based resolution.   However, it appears that some of the narrow-line emission does indeed stem from SN1957D itself.  We attribute the fact that the integrated \oiii\ flux values (Table  3) are lower than the those from imaging (Table 1) to the 1\farcs25 slit's having excluded a portion of the SN1957D flux, and/or to a slight mis-positioning of the slit.  The flux values from spectroscopy shown in Fig.{\ref{fig_time} are systematically lower than those from imaging at similar epochs, probably for the same reason.

The broad-line profile appears to be non-Gaussian, probably reflecting the distribution of shocked ejecta within the young remnant.  Fig.~\ref{fig_multi_epoch} shows the  \oiii\ lines at various epochs from 1989 through 2011.  
In addition to the 2011 GMOS spectrum, these include spectra from epochs 1989 and 1991 reported by \citet{long92}, and a previously unreported spectrum obtained from the 4m Blanco telescope at CTIO in 1998 June.  
The broad line profile appears qualitatively rather similar at all  these epochs, and indeed at all epochs since SN1957D's optical recovery \citep{long89, turatto89, cappallaro95, milisavljevic12}.  
 The combined flux in the broad component of the \OiiiL\ lines in 2011 is $4.2 \times 10^{-16} \FLUX$; all of the broad and narrow line fluxes are given in Table~\ref{table_spectrum}.

\subsection{The Stellar Population near SN1957D}

The stars in the region of M83 that contains SN1957D are very blue, indicative of a young association (see Fig.\   \ref{fig_overview} and the expanded view in Fig.\ \ref{fig_inset}). There is a fair amount of dust in the area, but the very blue colors of most of the stars suggest that the association is mostly on a fairly clear line of sight with minimal extinction.  There are some red giants in the general vicinity (the bright orange stars in Fig.\ \ref{fig_inset}), but none is within 2\farcs2 (50 pc) of the SNR and none is within the concentration of stars that defines the association. The absence of red giants suggests a relatively young age for the association.  To quantify this, we carried out DAOPHOT aperture photometry of stars in the association, as defined by a rectangular box  with dimensions $52 \times 35$ pixels ($2\farcs1 \times 1\farcs4$ or $46 \times 31$  pc), with the SNR near the right side of the rectangle (i.e., the upper left two-thirds of the field-of-view in Fig.\ \ref{fig_inset}).   We used an aperture radius of 3 pixels (0\farcs12), background sky annulus from 10 to 13 pixels, and the standard encircled-energy expression to account for the missing flux outside the 3 pixel aperture.

We corrected the magnitudes for Galactic foreground extinction using a value of $A_V$  = 0.21,\footnote{The average of estimates of the Galactic extinction by \cite{burstein82} and \cite{schlegel98}.}   and produced a color-color diagram shown in the left panel in Fig.\  \ref{fig_sed}.  For comparison, we also plot the Padova stellar isochrones for solar metallicity stars of age 4 and 10 Myr \citep{bertelli94, bertelli09}. The stars in the M83 stellar association are near the blue tip of the stellar isochrones, nearly coincident in the color-color diagram, as appropriate for very young stars.  There  is more scatter in the $V-I$ direction
than the $U-B$ direction since these are very blue stars (i.e., some are quite faint in the I band). Only stars brighter than $M_V$  = -4 are included in our analysis. The median value of $V-I$ is  -0.2, which matches the models very well, suggests that there is minimal internal extinction for most of the stars. However, the six reddest stars (solid square symbols in Fig.\  \ref{fig_sed}) do appear to be aligned roughly along the vector expected from reddening (dashed line). Hence we have corrected these stars individually for extinction using the technique described in \cite{kim12}. The corrected data are shown by the open squares. 

The right panel of Fig.\  \ref{fig_sed} shows the corresponding color-magnitude diagram for the stars in the SN1957D association, using the same symbol definitions as in the left panel. The locations in the diagrams are shown both uncorrected and corrected for internal reddening. After the data for the six reddest stars have been corrected for reddening, essentially all of the data are consistent with the stars being on the main sequence of a 4 Myr isochrone. A 10 Myr isochrone is also included, representing the oldest possible age for the association (i.e., the brightest star would have transitioned to a red giant).
The turnoff masses  
for the 4 and 10 Myr isochrons are 38 and 17 \MSOL\  respectively, both consistent with that expected for a core-collapse SN.

\section{Discussion \label{sec_discussion}}

SN1957D is not the only decades-old SN (or very young SNR) to have detectable X-ray emission. As a result of the increasing sensitivity of X-ray telescopes, primarily \chandra\ and XMM-{\em Newton}, about a dozen old SNe have been detected as X-ray sources a decade or more after the SN explosion \citep{soria08,dwarkadas12}. The oldest extragalactic SN that has been detected in X-rays today is SN1941C at a luminosity of \EXPU{5}{37}{\LUM} \citep{soria08}.  At age 24 years, SN1987A  has an X-ray luminosity of about \EXPU{4}{36}{\LUM}, though unlike SN1957D, its spectrum is dominated by soft X-rays in the range 0.3-2 keV that arise from a shock-heated thermal plasma \cite[see, e.\ g.][and references therein]{park11}.  Others range in X-ray luminosity from about \POW{37}{\LUM} to \POW{41}{\LUM}.  Most of these have X-ray spectra that are soft, and in cases where the sources are sufficiently bright, such as SN1979C in M100 \citep{immler05,patnaude11} and SN1993J in M81 \citep{chandra09}, evidence of emission from lines.  The fluxes are usually constant or declining with time \citep{dwarkadas12}. Their X-ray properties are normally interpreted as the interaction of the SN shock with circumstellar material, although \cite{patnaude11} have suggested that SN1979C has a non-thermal component as well.  All are assumed to be core-collapse SNe. The lower limit of about \POW{37}{\LUM} for a detected X-ray SNR reflects, for the most part, the sensitivity limits of the existing observations given the distances of nearby galaxies.   X-ray emission has not been reliably detected from any Type Ia SN in this age range \citep[see, e.g.][]{hughes07}.

The X-ray spectrum of SN1957D is featureless and hard, unlike the other SNRs in this age range that have been observed in X-rays and unlike the majority of Galactic SNRs.  Objects that do resemble SN1957D are pulsar-powered SNRs such as the Crab Nebula and the SNR 0540-69.3 in the Large Magellanic Cloud. The Crab Nebula is 958 years old; SNR 0540-69.3 is thought to be between 700 years old from the expansion of the filaments \citep{williams08} and 1700 years old from the pulsar spin-down age \citep{gradari11}.  X-rays from both are dominated by non-thermal emission with photon indices of 2.1 \citep{weisskopf10} and 2.0 \citep{hirayama02}, respectively, compared to SN1957D with $1.4^{+1.0}_{-0.8}$ .  The 0.3-8 keV luminosities of all three objects are about the same, $\sim$ \EXPU{1-2}{37}{\LUM}.   Thus, it is  plausible that X-ray emission in SN1957D is powered by a pulsar.  

The column density that we derive from the X-ray spectrum of SN1957D is quite high, \EXPU{2.0^{+1.6}_{-1.1}}{22}{cm^{-2}}, much higher than expected from foreground absorption in the Galaxy or in M83.  As discussed by \cite{fransson87} for SN1987A, cold ejecta from a core-collapse SN explosion provide substantial opacity for many years after the explosion.  Indeed, if the event that generated SN1957D  is similar to that used by  \cite{fransson87} for modeling SN1987A, one would predict that today the energy with optical depth unity would be roughly 5-6 keV.  Thus, if X-ray emission in SN1957D is powered by a pulsar it is not surprising that the source would be highly absorbed.  

The points marked with an X in the color-color and color-magnitude diagrams shown  Fig.\  \ref{fig_sed} represent the star that is coincident with
the SNR position.    With a reddening A$_v$ of about 2.1, this star is  the reddest of the stars in the association.  The other five stars we dereddened to place on the stellar isochrones have an average A$_V$ of 1.4, with a maximum value of about 1.8.  As mentioned earlier,  SN1957D is on the edge of the association and local extinction variations could account for the higher reddening of the star.   A more exotic idea would be that this is the companion to the SN and that localized dust formed in the SN ejecta is affecting the reddening.  A reddening of  A$_V\sim$2.1 corresponds to a column density of \EXPU{3.3}{21}{cm^{-2}} if the gas properties are typical of the ISM \cite{savage79}.  This is far less than the column density, \EXPU{2.0^{+1.6}_{-1.1}}{22}{cm^{-2}}, we derive for the X-ray source, but  there is no reason to believe that the properties of the gas in SN ejecta resemble that of the ISM.  Indeed the dust content of young SNe have turned out to be quite low compared to predictions based on normal gas-to-dust ratios \cite[see, e.g.][]{nozawa10}.  As a result, we have no observational way to distinguish between the possibility that the star is physically associated with the SN, or that its higher reddening is due to local extinction variations.


Anomalous X-ray pulsars (AXPs), thought to be ultramagnetized (B\EXPU{> 3}{14}{G}) isolated neutron stars \cite[see, e.g.][for a review of their properties]{mereghetti08}, are in principle another way to power emission from young SNRs .  They are known to be associated with young SNRs, and, since normal pulsars, with lower magnetic fields, generate pulsar wind nebulae, one suspects that they should be able to produce pulsar wind nebulae as well.  However, the average luminosity of AXPs is typically \POW{34-36}{\LUM},  less than is needed to power the observed emission from the remnant of SN 1957D and AXP X-ray spectra usually show steeper spectral indices than we observe from the remnant of SN1957D.  Furthermore, at present, the existence of (even faint) pulsar wind nebulae around AXPs is controversial \citep{younes12}. While we cannot rule out an AXP explanation for what we observe in SN1957D, given the state of our knowledge of AXPs, there is no need to invoke this explanation. The only obvious way to distinguish between a normal pulsar and an AXP would be to detect it directly, most likely by observing a soft $\gamma$ ray outburst in the direction of SN1957D.  This is not possible with today's instrumentation.  In the absence evidence for an AXP in SN1957D, a more normal pulsar/pulsar wind nebula, similar to the one that exists in the Crab Nebula is favored.


\cite{patnaude11}, in order to explain a non-thermal component in the X-ray spectrum of SN1979C, discuss the possibility that the emission might arise from a black  hole accreting material from a SN fall-back disk or from a binary companion.  The X-ray luminosity of SN1979C is  \EXPU{6.5}{38}{\LUM},  comparable to the Eddington luminosity for a stellar BH, which was one of their reasons for arguing that this system might require a black hole.  SN1957D is 30 times less bright in X-rays and so a rapidly spinning pulsar of the type known to exist in the Crab Nebula and  SNR 0540-69.3 can provide the needed luminosity for thousands of years.  A black hole origin for the X-ray emission in SN957D cannot be ruled out of course, but again, there is no indication that it is required in SN1957D, and therefore a more normal pulsar wind/nebula is favored for the X-ray emission.


In contrast to its X-ray properties, the radio and optical properties of SN1957D are not unusual for very young core-collapse SNRs.  Radio emission in young core-collapse SNRs is generally interpreted as synchrotron emission generated as the primary SN shock interacts with the circumstellar medium  \citep{chevalier82}.   The radio luminosity of SN1957D is not exceptional.  A secular decline in radio flux for a core-collapse SN as $F_{\nu} \propto t^{-0.7}$ is expected if the circumstellar medium is the stellar wind from a supergiant progenitor, where  the density scales as $r^{-2}$ \citep{chevalier82, weiler02}. The decline should steepen when the blast wave overruns the edge of the circumstellar material, and then flatten as it enters the undisturbed ISM.  
Breaks in the radio light curve have been measured for the recent type IIb SNe 1993J and 2001gd \citep{weiler07, stockdale07}.  In both cases, the breaks are  attributed to a decrease in the density of the circumstellar medium associated with a sudden increase in the mass-loss rate in the years prior to the explosion. The abrupt change in the radio flux is particularly well documented for SN1993J, beginning 8.5 yrs after the explosion. This appears to be what has happened in the case of SN1957D.  Between 1984 and 1990, it declined roughly at the canonical rate \citep{cowan94}. However based on our new ATCA data, the rate has been declining as  $t^{-4}$ since about 1990.  This suggests,  assuming a wind speed of 10$\VEL$ and a  velocity for the shock of 15,000$\VEL$ \citep{weiler07}, that the increase in mass loss from the progenitor of  SN1957D occurred  about 8000 yrs prior to explosion.   

Similarly, as noted in Section \ref{sec_intro}, the optical spectra of young core collapse SNRs are often but not always dominated by emission from forbidden lines of oxygen \citep{milisavljevic12}.   The optical emission from young core-collapse  SNRs, whether dominated by oxygen lines or not, is understood to arise from shock interactions within the ejecta and between the forward shock and the circumstellar medium \citep{chevalier94}.  For the oxygen-dominated remnants, such as SN1957D,  the best picture to date  one that involves the  interaction between fast-moving, dense oxygen-dominated fragments from the explosion and less dense, perhaps more evenly distributed SN ejecta of the reverse shock \cite{southerland95}.  (This reproduces the observed line ratios better than simpler models involving the reverse shock propagating back into the ejecta.) The interaction between the fragments and the more tenuous material creates a bow-shock around the fragments, or knots, and drives shocks into the fragments themselves.  Optical emission arises from the shocked oxygen-rich ejecta in the knots.   Optical spectra of SN1957D are similar to those of other O-rich SNRs, including Cas A, G292.0+1.8 and E0102-72.3, that lack luminous pulsar/pulsar wind nebulae.
 
Nevertheless, in view of the unusual X-ray spectrum of SN1957D and the suggestion that its X-ray emission arises from a pulsar/pulsar wind nebula, it is worthwhile asking whether the radio or optical emission could be due to a pulsar as well.

The radio flux from the pulsar wind nebula within SNR 0540-69.3  is 520 mJy at 5 GHz \citep{manchester93}.  Scaled to the distance of M83, this flux would be 61 $\mu$Jy, compared to the 230 $\mu$Jy at 5 GHz that we measure from SN1957D.  Using parameters given by \cite{green09}, the Crab Nebula would have a flux of about 120 $\mu$Jy at 5 GHz at the distance of M83, about half that of SN1957D in the latest ATCA observations. Thus, while the radio luminosity of SN1957D is larger than either of these two objects, it is not so large as to make implausible the possibility that the radio emission is powered by a pulsar (if we include the pulsar wind nebula in the estimate of the luminosity).  However, a problem with the hypothesis that a pulsar dominates at radio wavelengths is how to reconcile in a natural manner the time evolution of the flux  at radio and X-ray wavelengths.  We know, based on the upper limits to X-ray flux from the 2000-2001 \chandra\ observations, the X-ray luminosity in 2000-2001 was about the same as or possibly lower than  in 2011.  Although synchrotron lifetimes for electrons differ at different wavelengths, very special circumstances would appear to be required to create a rapidly declining flux at radio wavelengths and a near constant (or increasing) flux at X-ray wavelengths.

The optical lines generated in pulsar-dominated supernova remnants, such as the Crab Nebula,  are produced primarily by photoionization of ejecta (and the ISM) by synchrotron emission.  \cite{chevalier92} discussed the development of UV-optical-IR emission lines in pulsar-dominated core-collapse SNR.  In very young SNRs, the lines are narrow and permitted lines are favored as the ionization progresses into the expanding ejecta, but the velocity widths broaden with time and forbidden lines, especially [O III] become more prominent as the density drops.  
The optical spectra of the Crab Nebula and  SNR 0540-69.3 are similar to that observed in SN1957D, with broad emission from the lines of oxygen and sulfur \citep[see, Fig.\  6 of][for a comparison]{morse06}, although \HA+\NiiL\ is much more obvious in the integrated optical spectrum of the Crab.   The \OiiiL\ emission-line fluxes in SN1957D have been nearly constant since 2002, which is consistent with the X-ray fluxes, but if we speculate that the \OiiiL\ flux scales with emission from the pulsar, then the pulsar/pulsar wind nebula must have been dramatically brighter prior to 1990, which is not easily understandable.     Moreover, in the \cite{chevalier92} models, \OiiiL\ should brighten with time and the line widths should broaden with time.   That is not what is observed in SN1957D.

A final possibility is that  the radio emission, optical emission, and the X-ray emission all arise from the shocks between ejecta and circumstellar material. But in this case, one would again need to explain the hard, featureless spectrum and the fact that the X-ray object was not brighter in the 2000-2001 \chandra\ observations.    Unfortunately, it is unlikely that a better X-ray spectrum will be acquired any time in the near future, though it should be possible to monitor the X-ray flux. Possibly the X-ray (and radio emission) arise  primarily from  synchrotron emission at the shock front.  However, to our knowledge, none of the SNRs of this type is nearly as bright as SN1957D in X-rays and none has emission lines arising from ejecta.  Perhaps the most interesting of these objects is the Milky Way SNR G1.9+0.3, which, with an estimated age of 140$\pm$30 years, is the youngest known SNR in the Galaxy.  It has a very hard non-thermal spectrum, similar to that observed in SN1957D, but its X-ray luminosity is $\sim$ \EXPU{3}{34}{\LUM}, about 500 times fainter than SN1957D, consistent with its being the remnant of a Type Ia explosion \citep{reynolds09}.  The alternative, emission from a very hot thermal plasma, also seems hard to justify, given the X-ray spectra of other young core-collapse SNe.

In conclusion, while we caution the reader to keep in mind the fact that the X-ray spectrum contains only 140 counts, the existing data favor a picture of SN1957D in which the X-ray emission arises from a pulsar/pulsar-wind nebula, while the optical and radio emission arise primarily in shocks in the ejecta and circumstellar gas.  In this picture, X-rays are produced in a region within the cloud of cold ejecta and are thus heavily absorbed. The optical emission arises in the vicinity of the reverse shock, but outside the cold, freely expanding ejecta. The radio emission is produced at the shock front.

\section{Summary \label{sec_summary}}

We have detected X-ray emission from the remnant of SN1957D for the first time using a very deep \chandra\ observation of M83.  The X-ray spectrum of the source is unusually hard and has unusually high column density for a young  SNR.  The X-ray spectra suggest that the SNR is dominated by an X-ray pulsar with substantial absorption in the ejecta.  The SN was almost certainly the core-collapse explosion of a massive progenitor, based on its association with a young star cluster with an age of 4-10 million years,  the high star formation rate in the spiral arms of M83 generally.  At this age, the turnoff mass is at least 17\MSOL, exactly where one would expect a core-collapse SN.  Optically, the emission is dominated by \OiiiL, suggesting a shock interacting with ejecta that is highly enriched in oxygen.  The shape of the optical spectrum has not evolved significantly since the first spectrum was obtained in 1987, 30 years after the explosion, but the flux has declined at both optical and radio wavelengths indicating that the density of the medium into which the shock is expanding has also declined.  Given the unusual nature of the X-ray spectrum, it is important to continue to monitor the evolution of this very young SNR at all wavelengths.

\acknowledgements

Support for this work was provided by the National Aeronautics and Space Administration 
through \chandra\ Grant Number G01-12115, issued by the \chandra\ X-ray Observatory Center, 
which is operated by the Smithsonian Astrophysical Observatory 
for and on behalf of NASA under contract NAS8-03060\@.   
PFW and WPB are grateful for both observing and travel support 
for the Gemini observations from the Gemini office at NOAO.  
PFW also  acknowledges financial support from the National Science Foundation 
through grant AST-0908566, 
and the hospitality of the Research School of Astronomy and Astrophysics, 
Australian National University, during a portion of the work presented here.

The Gemini Observatory is operated by Association of Universities for Research in Astronomy, Inc., under a cooperative agreement
with the NSF on behalf of the Gemini partnership: the National Science Foundation (United
States), the Science and Technology Facilities Council (United Kingdom), the
National Research Council (Canada), CONICYT (Chile), the Australian Research Council (Australia),
Minist\'{e}rio da Ci\^{e}ncia, Tecnologia e Inova\c{c}\~{a}o (Brazil)
and Ministerio de Ciencia, Tecnologia e Innovaci\'{o}n Productiva (Argentina).

\clearpage

\begin{deluxetable}{ccccccccc}
\tabletypesize{\scriptsize}
\rotate
\tablewidth{0pt}
\tablecaption{SN1957D Imaging Observations and Fluxes}

\tablehead{
\colhead {} &\colhead {} & \colhead{} &
\multicolumn{3}{c}{Filter} & 
\colhead {} & 
\colhead {} &
\colhead{[O\,III] 5007\,\AA\ Flux}\\ 

\cline{4-6}  

\colhead{Date (U.T.)} & 
\colhead {Telescope} &
\colhead {CCD} &
\colhead{Line} &
\colhead{$\lambda_c\;$(\AA)} &
\colhead{$\Delta \lambda$\tablenotemark{a}(\AA)} &
\colhead {Exposure (s)} &
\colhead {Observers} &
\colhead{$10^{-16}\FLUX$}
}

\startdata
1987 Apr 26\tablenotemark{a} &LCO 2.5 m & TI  & [O III] & 5020 &54\phn\phn & $2\times1000$\phn\phn & KSL, WPB & 120\\
&&&Continuum& 4770 & 100\phn\phn&&W. Krzeminski&\\
&&&&&&&&\\
1991 Apr 18\tablenotemark{b} & CTIO 4.0 m &TEK 1K \#1&[O III]&5020&54\phn\phn&$3\times600$\phn\phn&PFW, KSL & 24 \\
&&&Continuum&4770&100\phn\phn&3$\times$500\phn\phn& &\\
&&&&&&&&\\
2002 Mar 20-23&CTIO 0.9 m&TEK 2K \#3 &[O III]&5006&60\phn\phn&$4 \times 1000$\phn\phn&PFW, KSL, & $8.9 \pm 2.6$\\
&&&Continuum&5135&90\phn\phn&$4 \times 500$\phn\phn&C. Reith &\\
&&&&&&&&\\
2007 Apr 15 & SOAR 4.1 m & SOI & [O III] & 5006 & 60\phn\phn & $ 4\times 600$\phn\phn & PFW, KSL & $8.1\pm 1.6$ \\
&&&Continuum&5316&161\phn\phn&$7 \times 200$\phn\phn & &\\
&&&&&&&&\\
2009 Apr 26 & Magellan I 6.5 m & IMACS & [O III] & 5007 & 50\phn\phn & $ 7\times 600$\phn\phn & PFW, KSL & $8.0\pm 1.0$ \\
&&&Continuum&5316&161\phn\phn&$7 \times 200$\phn\phn & &\\
&&&&&&&&\\
2010 Mar 19 & \hst & WFC3 & [O III] & 5010 & 65\phn\phn & $3\times 828$\phn\phn &R. O'Connell & $8.9 \pm 0.5$ \\
&&&&&&&&\\
\enddata
\label{table_optical}
\tablenotetext{a}{\citet{long89}}
\tablenotetext{b}{\citet{long92}}

\end{deluxetable}

\begin{deluxetable}{ccccc}
\tablecaption{HST Image Summary$^a$ }
\tablehead{\colhead{Band} & 
 \colhead{Association~ID} & 
 \colhead{Filter} & 
 \colhead{Exposures} & 
 \colhead{Total~Exposure} 
\\
\colhead{~} & 
 \colhead{~} & 
 \colhead{~} & 
 \colhead{~} & 
 \colhead{(s)} 
}
\tabletypesize{\scriptsize}
\tablewidth{0pt}\startdata
U &  IB6WB1060 &  F336W &  3 &  2560 \\ 
B &  IB6WB1050 &  F438W &  3 &  1800 \\ 
V &  IB6WB2070 &  F547M &  4 &  1203 \\ 
I &  IB6WB2060 &  F814W &  3 &  1213 \\ 
{[OIII]} &  IB6WB2050 &  F502N &  3 &  2484 \\ 
H$\alpha$ &  IB6WB2010 &  F657N &  4 &  1484 \\ 
{[SII]} &  IB6WB3030 &  F673N &  3 &  1770 \\ 
\tablenotetext{a}{ Images obtained in 2010 March as part of Early Release Observations for WFC3. Program ID: 11360, PI: Robert O'Connell}
\enddata 
\label{table_hst}
\end{deluxetable}

\begin{deluxetable}{ccr}
\tabletypesize{\small}
\tablewidth{0pt}
\tablecaption{SN1957D Emission-line Fluxes, 2011 April}

\tablehead{

\colhead{} & \colhead{} &
\colhead{Flux\tablenotemark{b}} \\
\colhead{Ion} &
\colhead{$\lambda_0$\,(\AA)\tablenotemark{a}} &
 \colhead{$10^{-17}\FLUX$} \\
 \\
\multicolumn{3}{c}{Broad Lines, $\Delta \lambda = 2730 \kms$}

}

\startdata

[O III] & 4959 & 12.1\phn\phn\phn\phn\phn\phn\phn \\

[O III] & 5007 & 30.2\phn\phn\phn\phn\phn\phn\phn \\

[O I] & 6300 & 10.1\phn\phn\phn\phn\phn\phn\phn \\

[O I] & 6363 & 3.0\phn\phn\phn\phn\phn\phn\phn \\

[S II] & 6716 & 8.5\phn\phn\phn\phn\phn\phn\phn \\

[S II] & 6731 & 5.9\phn\phn\phn\phn\phn\phn\phn \\

[Ar III] & 7135 & (5.7)\phn\phn\phn\phn\phn\phn \\

\hline
\\
\multicolumn{3}{c}{Narrow Lines (unresolved)} \\

\hline

H$\beta$ & 4861 &  14.8\phn\phn\phn\phn\phn\phn\phn \\

[O III] & 4959 & 4.7\phn\phn\phn\phn\phn\phn\phn \\

[O III] & 5007 & 11.8\phn\phn\phn\phn\phn\phn\phn \\

[He I] & 5875 & 3.2\phn\phn\phn\phn\phn\phn\phn \\

[O I] & 6300 & (1.0)\phn\phn\phn\phn\phn\phn\phn \\

[N II] & 6548 & 12.4\phn\phn\phn\phn\phn\phn\phn \\

H$\alpha$ & 6563 &  76.2\phn\phn\phn\phn\phn\phn\phn \\

[N II] & 6583 & 37.4\phn\phn\phn\phn\phn\phn\phn \\

[S II] & 6716 & 9.2\phn\phn\phn\phn\phn\phn\phn \\

[S II] & 6731 & 6.9\phn\phn\phn\phn\phn\phn\phn \\

[Ar III] & 7135 & 3.3\phn\phn\phn\phn\phn\phn\phn \\

\enddata

\label{table_spectrum}

\tablenotetext{a}{Rest wavelength}
\tablenotetext{b}{Values in parentheses are highly uncertain.}

\end{deluxetable}


\begin{figure}
\plotone{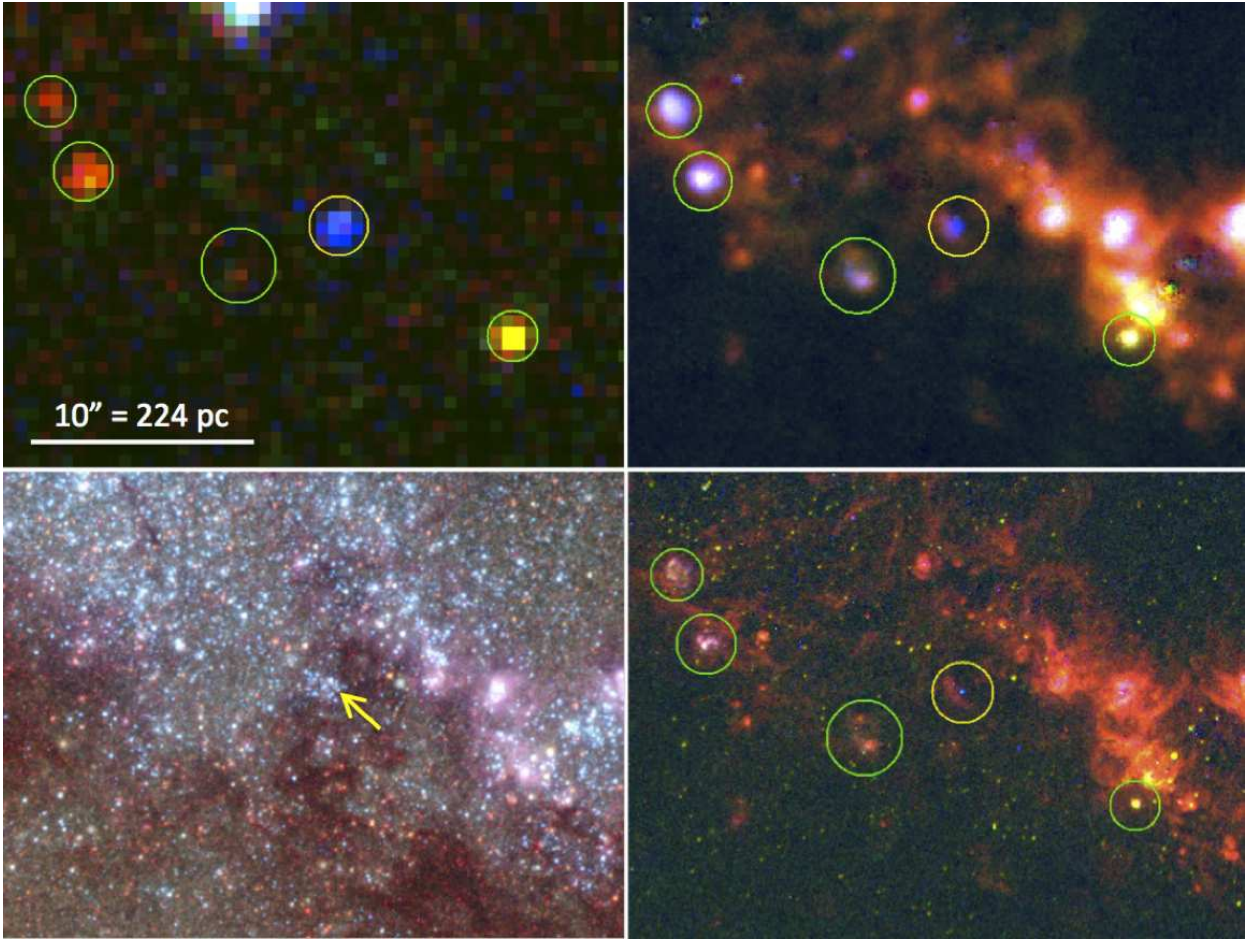}
\figcaption{A four panel context figure showing the region in the immediate
vicinity of SN1957D.  At upper left is a three-color \chandra\ image, where red is
0.35 - 1 keV, green is 1 - 2 keV, and blue is 2 - 8 keV.  At upper right is a
three-color Magellan optical continuum-subtracted emission line image, taken in exceptional 0\farcs4 seeing, where red
is H$\alpha$, green is \sii, and blue is \oiii.  At lower right is the same
as the upper right panel, but for the \hst\ WFC3 emission line data.  The lower
left panel is the \hst\ WFC3 continuum plus H$\alpha$ image of the same region,
provided by Zolt Levay (STScI), where red is F547M, green is B band, and blue is
U-band, with H$\alpha$ also shown as red/pink.  (Note the star subtraction on
the \sii\ data in the lower right panel is imperfect, leaving some stellar
residuals that can be used for reference in comparison with the lower left
panel.)   The green circles indicate normal ISM-dominated SNRs in the vicinity,
and the yellow circle indicates the SN1957D position.  The hard X-ray source
is positionally coincident with the optical SNR which is dominated by  \oiii;  the star indicated by  the arrow in the lower left panel is also coincident with the optical SNR (see also Fig.\ \ref{fig_inset} for an expanded view of this region).  North is up and east is
to the left, and a scale bar is shown in the upper left panel. \label{fig_overview}}
\end{figure}

\begin{figure}
\includegraphics[scale=0.6,angle=-90]{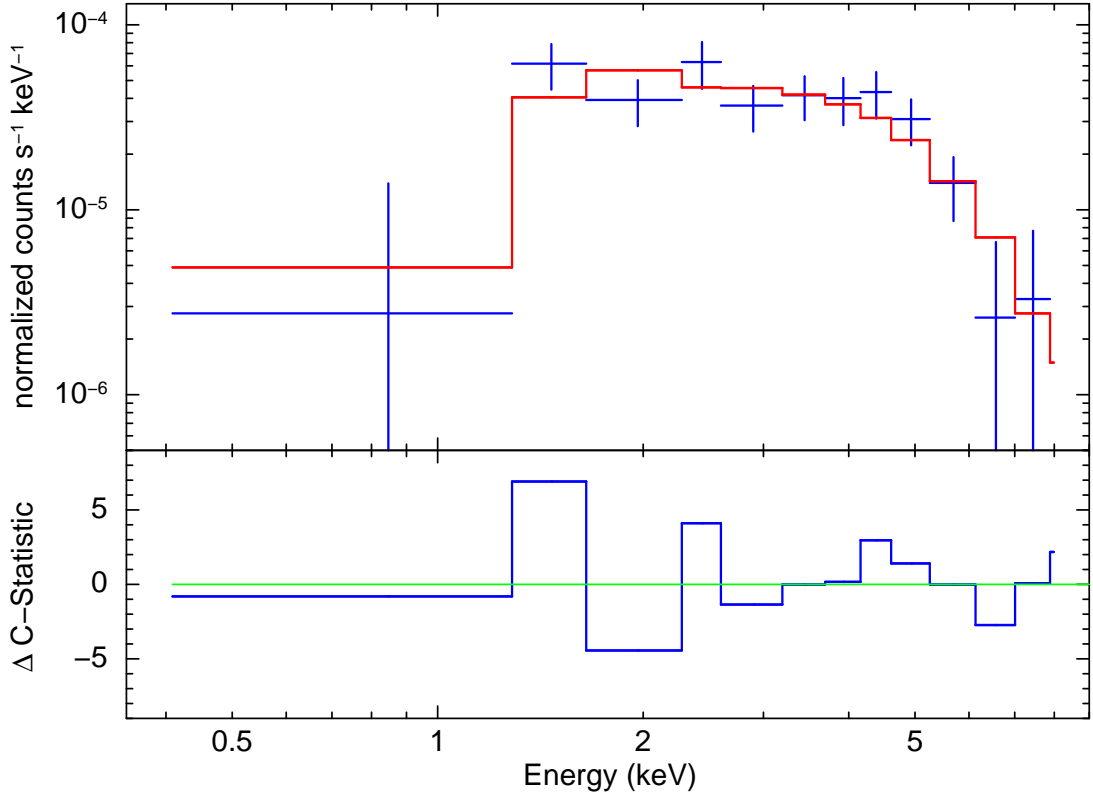}
\figcaption{X-ray spectrum of SN1957D fitted with an absorbed power law.   Datapoints and C-statistics residuals are shown.  After allowing for  Galactic absorption of \EXPU{4}{20}{cm^{-2}}, we find a photon power law index  $\gamma$ of $1.4^{+1.0}_{-0.8}$ and $N_H$(M83) of \EXPU{2.0^{+1.6}_{-1.1}}{22}{cm^{-2}}.
 \label{fig_xray_spectrum}}
\end{figure}

\begin{figure}
\epsscale{0.8}
\plotone{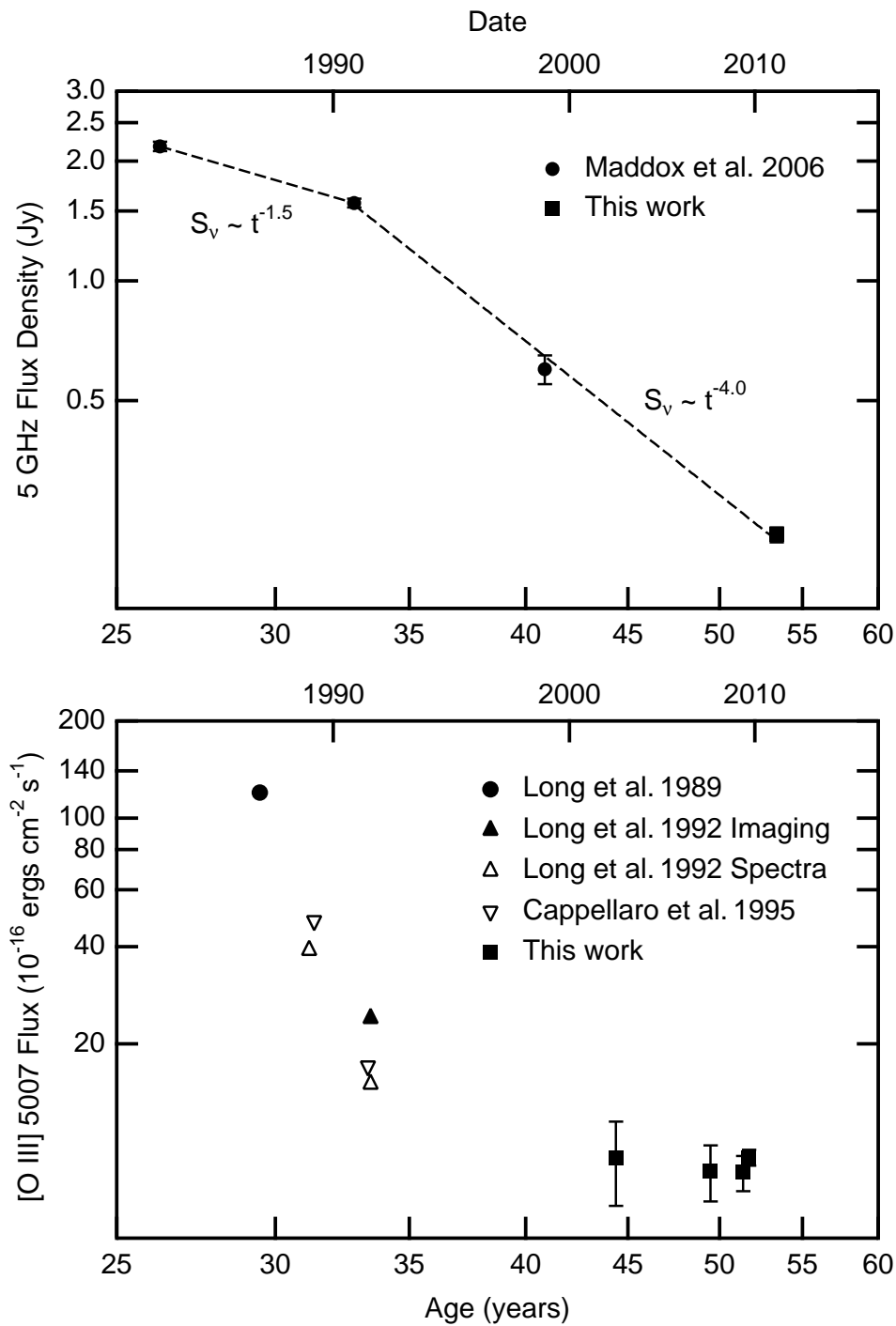}
\epsscale{1.0}

\figcaption{The time evolution of the radio flux (upper panel) and the optical \oiii\ $\lambda5007$ emission-line flux (lower panel) of SN1957D.   The horizontal axes are identical and the vertical axes have the same dynamic range in the two panels.  In the optical panel, filled symbols represent flux values from imaging; open symbols  represent values from spectra, which typically miss some flux.                                                                                                                                                                                                                          
\label{fig_time}}
\end{figure}

\begin{figure}
\plotone{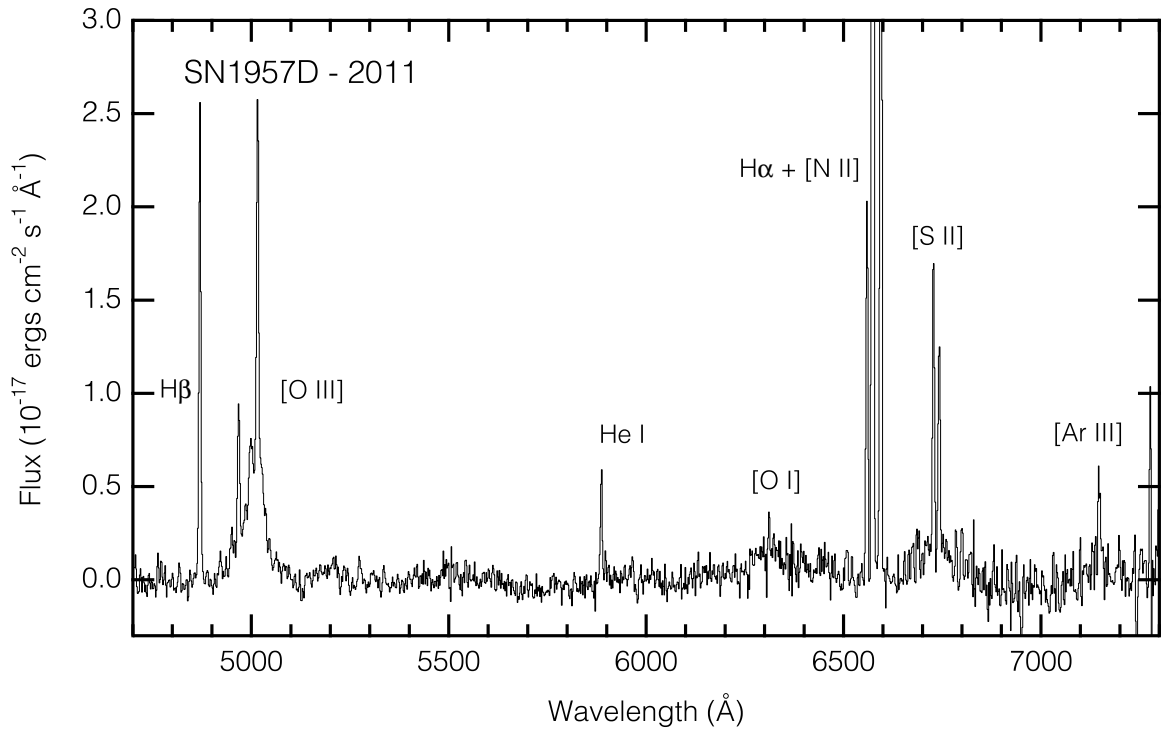}

\figcaption{Optical spectrum of SN1957D, as observed in 2011 April with GMOS on Gemini South.  The remnant shows broad lines from \OiiiL, \OiL, and \SiiL.  Narrow emission lines are also seen, but they most likely arise from photoionized gas seen to the east of the SNR in the \hst\ images. \label{fig_optical_spectrum}}
\end{figure}

\begin{figure}
\epsscale{0.8}
\plotone{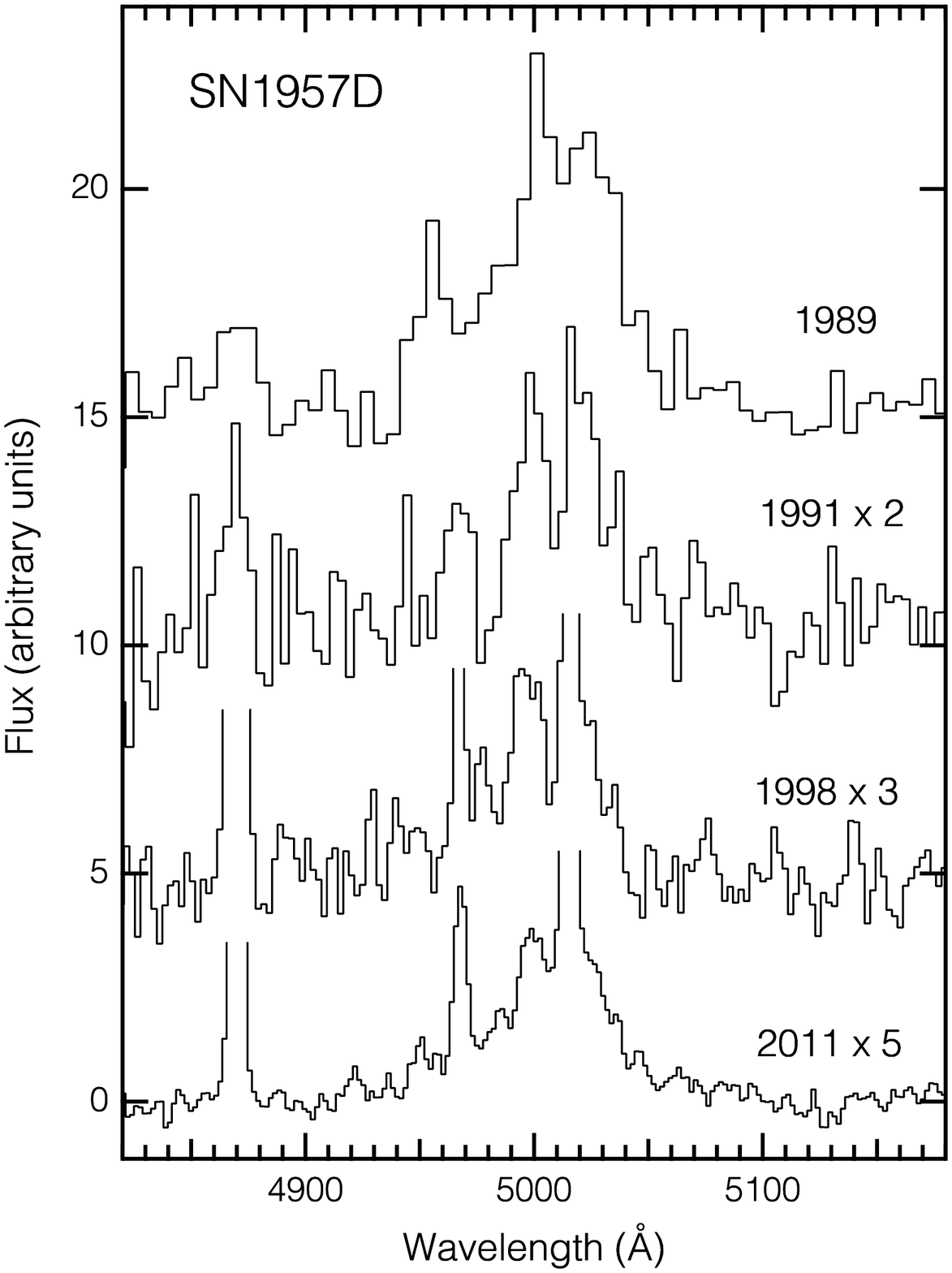}

\figcaption{
The emission-line profile in  \OiiiL\ region of the spectrum  as observed in 2011 April with GMOS (Fig.~\ref{fig_optical_spectrum}), and from 1989, 1991 \citep[both][]{long92} and from 1998 (previously unpublished CTIO Blanco data).  The broad line emission has decreased over time (note scaling), but the profile has remained qualitatively similar.  The narrow lines were largely removed from the 1989 and 1991 spectra, but have been retained (but truncated) in those from 1998 and 2011, where the higher resolution minimizes broad-narrow confusion.   \label{fig_multi_epoch}}
\end{figure}

\begin{figure}

\epsscale{1.0}
\plotone{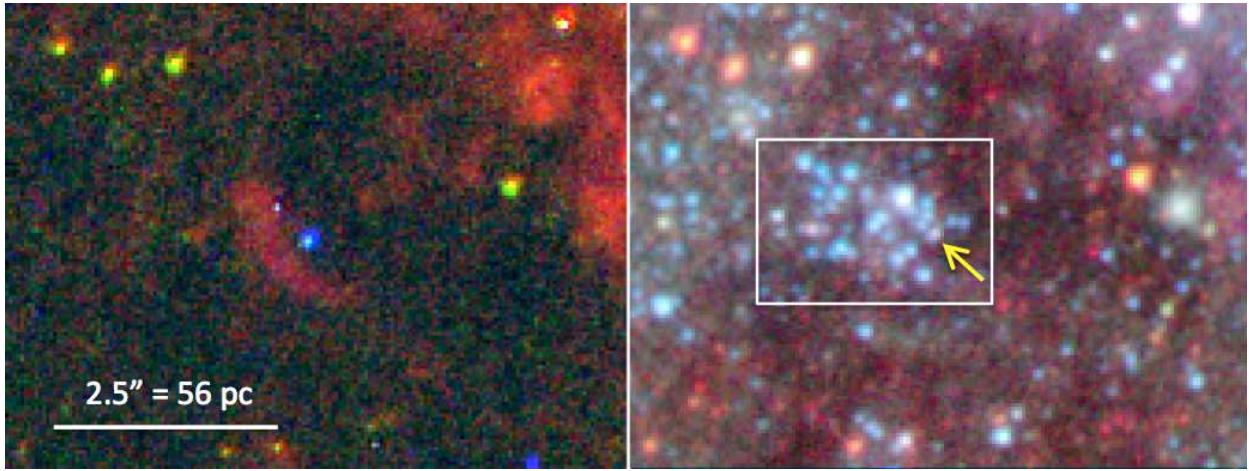}
\figcaption{The stellar association in which SN1957D is located as observed with \hst.  The left panel is a 3-color composite of the emission line gas, where red
is H$\alpha$, green is \sii, and blue is \oiii.  The right panel is a portion of the continuum plus H$\alpha$ image of the same region described more fully in the caption of Fig. \ref{fig_overview}. Stars within the rectangular shaped region were used to establish the age of the stellar association along the line of sight to SN1957D. \label{fig_inset}}
\end{figure}

\begin{figure}
\rotate
\includegraphics[scale=0.6,angle=-90]{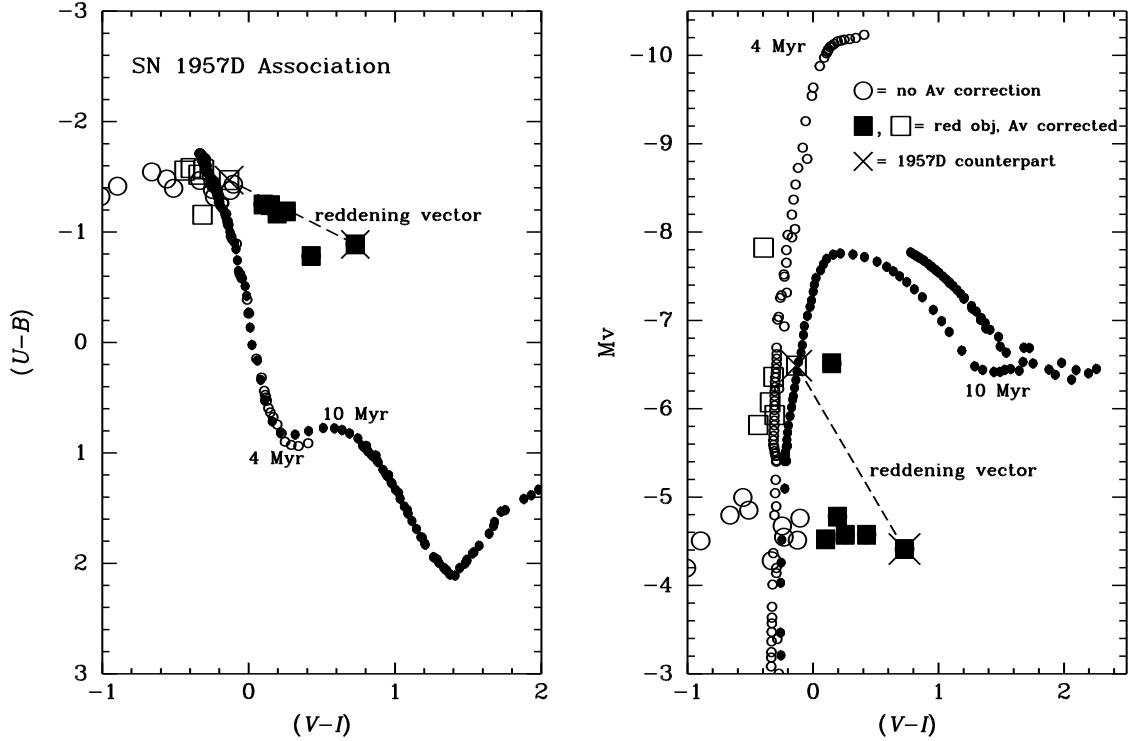}
\figcaption{Color-color (left panel) and color-magnitude (right panel) diagrams for stars in the stellar association along the line of sight to SN1957D.  The positions of the stars in the assocation are compared to stellar isochrones for solar metallicity stars of age 4 and 10 Myr \citep{bertelli94, bertelli09}.  For stars with significant reddening, both the uncorrected and reddening-corrected positions are shown. The stellar association is likely to be as young as 4 Myr and is almost certainly less than 10 Myr.  The X indicates the object coincident with SN1957D.
\label{fig_sed}}
\end{figure}

\end{document}